\documentclass[a4paper,11pt]{article}
\usepackage{pos_1}

\title{Elastic neutrino-atom scattering as a probe of neutrino millicharge and
magnetic moment}
\author[a]{Georgy Donchenko}
\author*[a]{Konstantin Kouzakov}
\author[a]{Alexander Studenikin}

\affiliation[a]{Faculty of Physics, Lomonosov Moscow State University,\\
  Leninskie gory, Moscow 119991, Russia}

\emailAdd{kouzakov@gmail.com}
\emailAdd{studenik@srd.sinp.msu.ru}

\abstract{Neutrino scattering on atomic systems at low-energy transfer is a powerful tool for searching the neutrino electromagnetic interactions. The regime of coherent elastic neutrino-atom scattering, i.e., when the atom recoils as a pointlike particle, can be effectively fulfilled in the case of tritium antineutrinos. We present theoretical calculations for coherent elastic neutrino-atom scattering processes on such targets as the H, $^2$H, $^3$He, and $^4$He %, and $^{12}$C
atoms. We show how the atomic effects and neutrino electromagnetic properties, namely the neutrino millicharge and magnetic moment, may manifest themselves in the atomic-recoil spectra. Our results can be used in planning the experiments on coherent elastic neutrino-atom scattering (in particular, with superfluid He-4).}

\FullConference{%
  *** The 41st International Conference on High Energy physics, ***\\
  *** 6-13 July 2022 ***\\
  *** Bologna, Italy ***
}

\begin{document}
\maketitle

\section{Introduction}
The search for light particles of dark matter requires the detectors that are sensitive to low recoil energies ($\lesssim$100 meV). This can be achieved, for example, by using a superfluid He-4 target \cite{Maris2017}.
Another possible application of the superfluid He-4 detector could be the study of the low-energy neutrino scattering, in particular of the coherent elastic neutrino-atom scattering (CE$\nu$AS)~\cite{Gaponov&Tikhonov,Sehgal&Wanninger,Cadeddu2019,Picciau2021} that has not been observed so far. Below we inspect the sensitivity of the CE$\nu$AS processes on light atomic systems to such neutrino electromagnetic properties as millicharge $e_\nu$ and magnetic moment $\mu_\nu$~\cite{RMP2015}. For this purpose we account for the indicated neutrino properties in the CE$\nu$AS cross section and present the corresponding numerical results.

\section{Effects of neutrino millicharge and magnetic moment in CE{\it v}AS}
We consider an elastic neutrino-atom collision in the following kinematical regime:
$$
E_\nu\ll m, \qquad T\leq\frac{2E_\nu^2}{m}\ll E_\nu, \qquad  E_\nu\ll\frac{1}{R_{\rm nuc}},
$$
where $E_\nu$ is the neutrino energy, $T$ is the energy transfer, $m$ is the atomic mass, and $R_{\rm nuc}$ is the nuclear radius.

According to \cite{Gaponov&Tikhonov,Sehgal&Wanninger,Cadeddu2019,Kouz2017,Donchenko2021,Donchenko2022}, the CE$\nu$AS differential cross section is given by
\begin{equation}
 \label{cr_sec}
            \frac{d\sigma}{dT}=\frac{d\sigma^{(w,e_\nu)}}{dT} + \frac{d\sigma^{(\mu_\nu)}}{dT}.
  \end{equation}
Here the weak interaction and neutrino millicharge contribution is
\begin{equation}
\frac{d\sigma^{(w,e_\nu)}}{dT}= \frac{G_F^2 m}{\pi}  \left[ C_V^2 \left(1 - \frac{mT}{2E_\nu^2} \right) + C_A^2 \left(1 + \frac{mT}{2E_\nu^2} \right) \right],
\end{equation}
with
    	\begin{align*}
    	C_V &= Z \left(\frac{1}{2} - 2 \sin^2 \theta_W \right) - \frac{1}{2} N + Z \left( \mp\frac{1}{2} + 2 \sin^2 \theta_W \right) F_{\rm el} ( q^2 )+\frac{\sqrt{2}\pi\alpha Ze_\nu}{G_FmT}[1-F_{\rm el} ( q^2 )],\\
        C_A^2 &= (C_A^{\rm nuc})^2 + \frac{1}{4}\sum\limits_{n,l} \left[\left( L^{nl}_{+} - L^{nl}_{-} \right) F_{\rm el}^{nl}(q^2)\right]^2,\\
(C_A^{\rm nuc})^2 &= \frac{g_A^2}{4} \left[ (Z_{+} - Z_{-} ) - (N_{+} - N_{-}) \right]^2,
    	\end{align*}
         where $q$ is the momentum transfer, with $q^2=2mT$, the plus (minus) stands for $\nu=\nu_e$ ($\nu=\nu_{\mu,\tau}$), and
        $Z$ ($N$) is the number of protons (neutrons) in the atomic nucleus. $F_{\rm el}(q^2)$ is the Fourier transform of the electron density, $g_A=1.25$, $Z_\pm$ and $N_\pm$ are the numbers of protons and neutrons (electrons) with spin parallel ($+$) or antiparallel ($-$) to the nucleus spin (the total electron spin). $L^{nl}_\pm$ is the number of electrons in the $nl$ atomic orbital with spin parallel ($+$) or antiparallel ($-$) to the electron spin, and $F_{\rm el}^{nl}(q^2)$ is the Fourier transform of the $nl$ electron density. The neutrino millicharge $e_\nu$ is in units of $e$.

%
%\section{Numerical results}
%
\begin{figure}%[htb]
    \centering
   \includegraphics[width=0.48\linewidth]{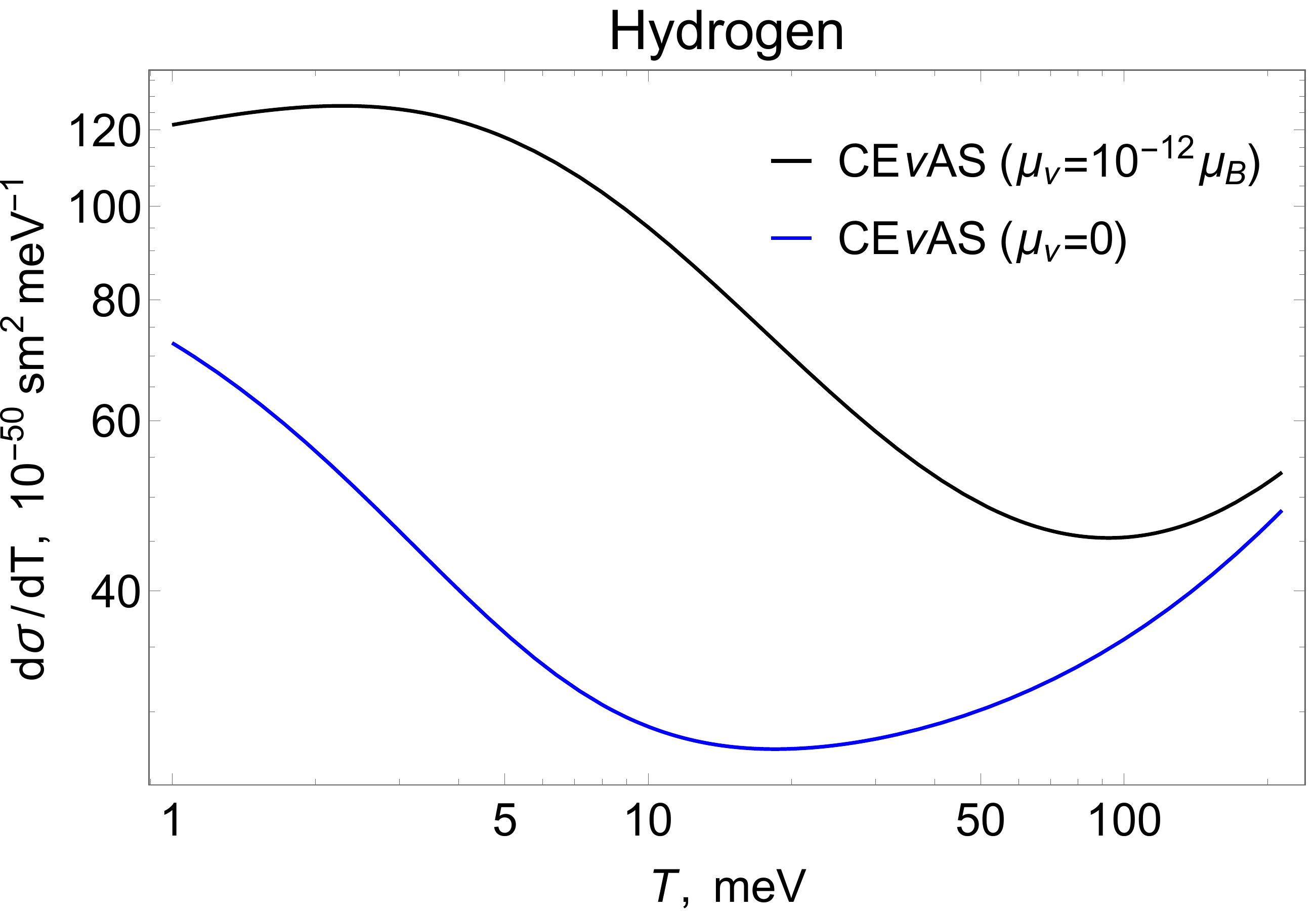}
    \hfill
    \includegraphics[width=0.48\linewidth]{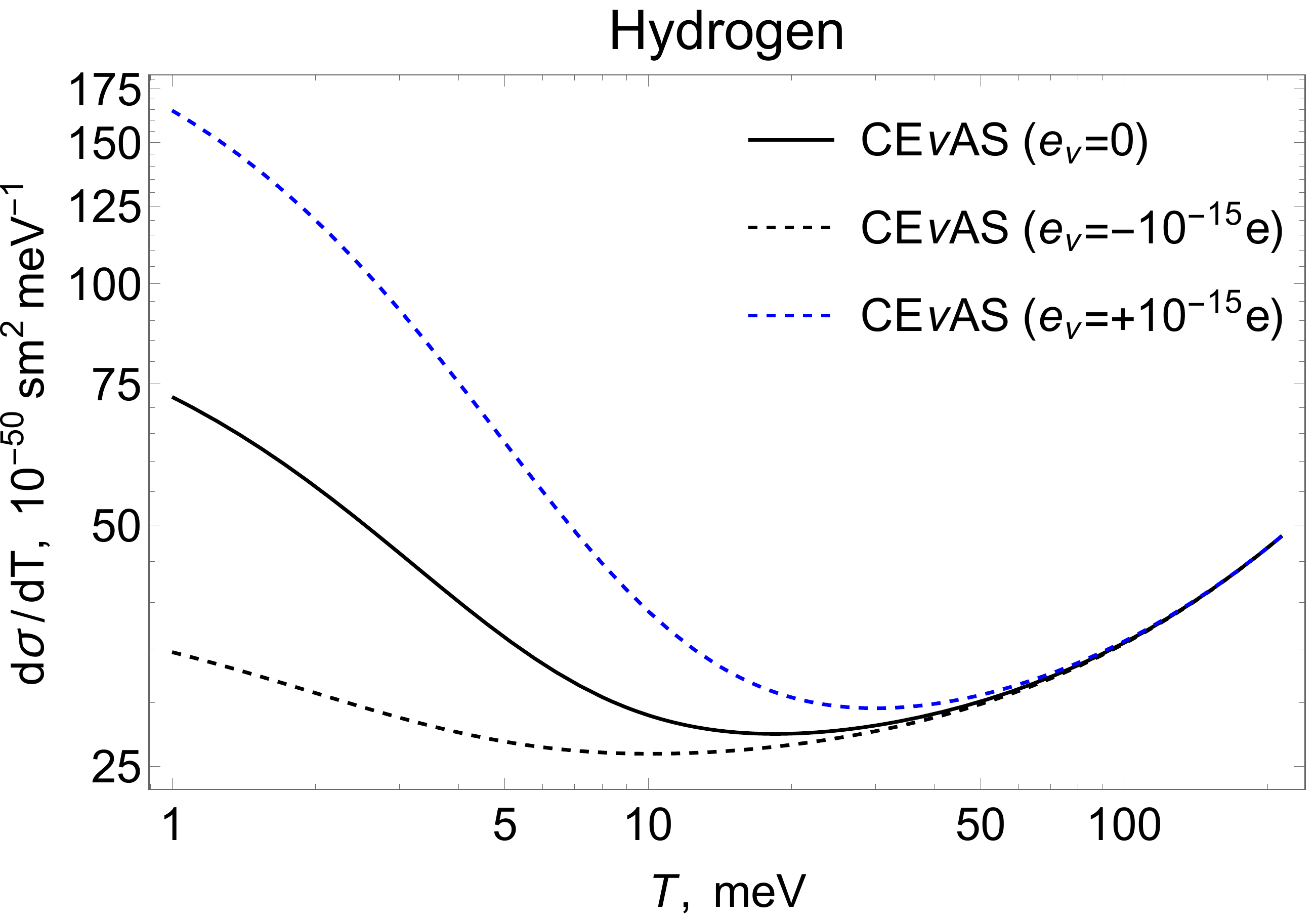} \\
   \vspace{3mm}
    \includegraphics[width=0.48\linewidth]{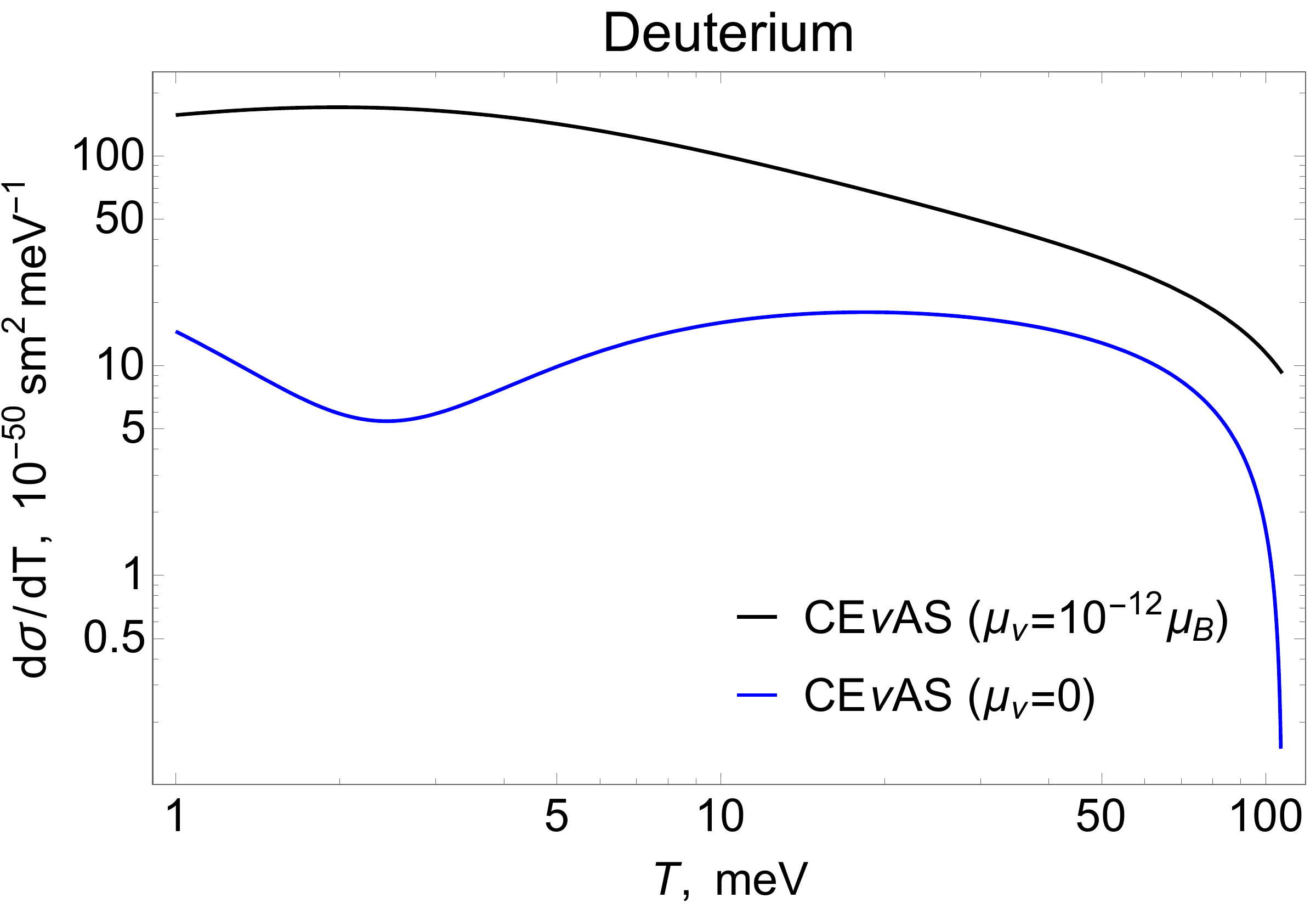}
    \hfill
    \includegraphics[width=0.48\linewidth]{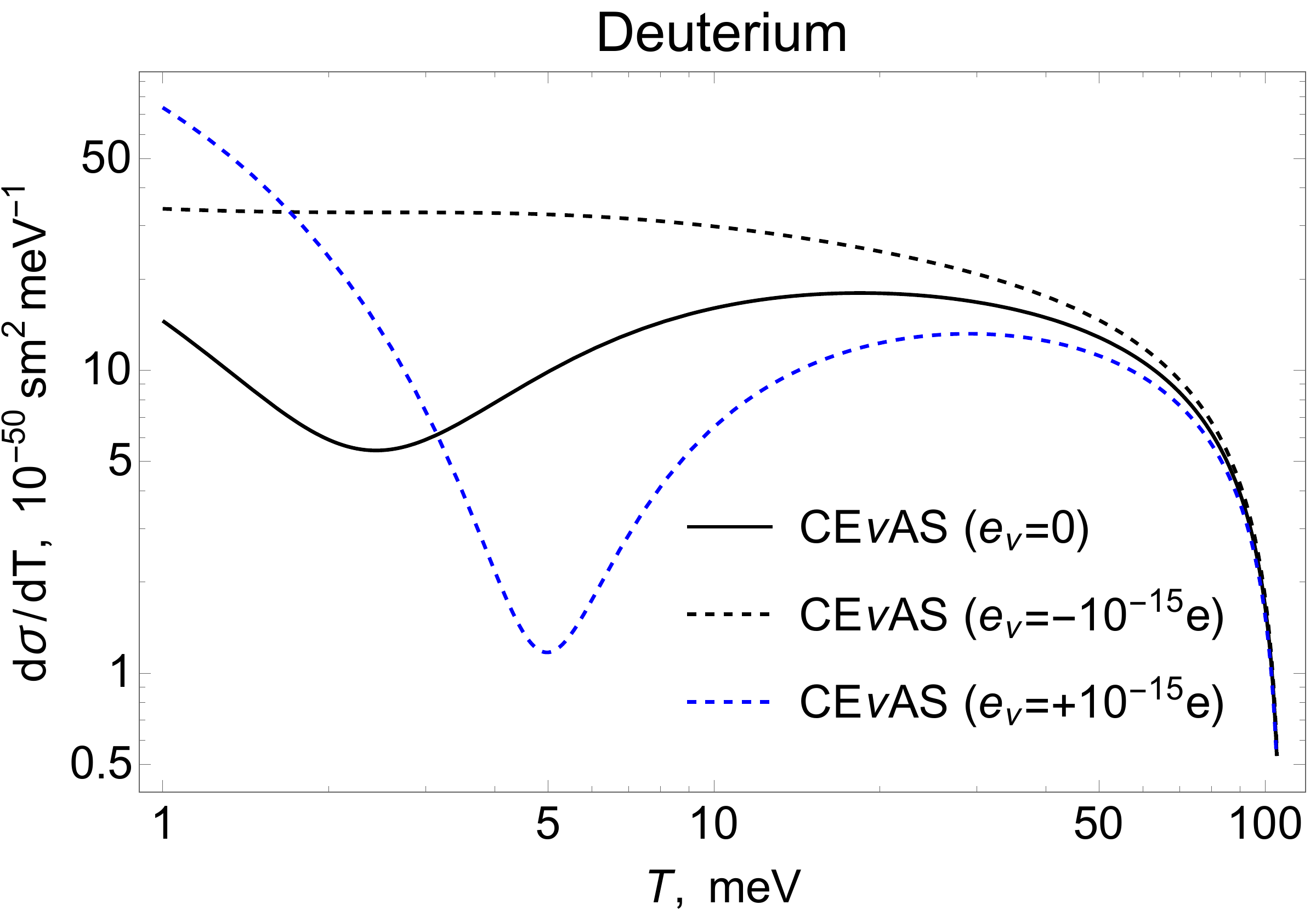}\\
   \vspace{3mm}
        \includegraphics[width=0.48\linewidth]{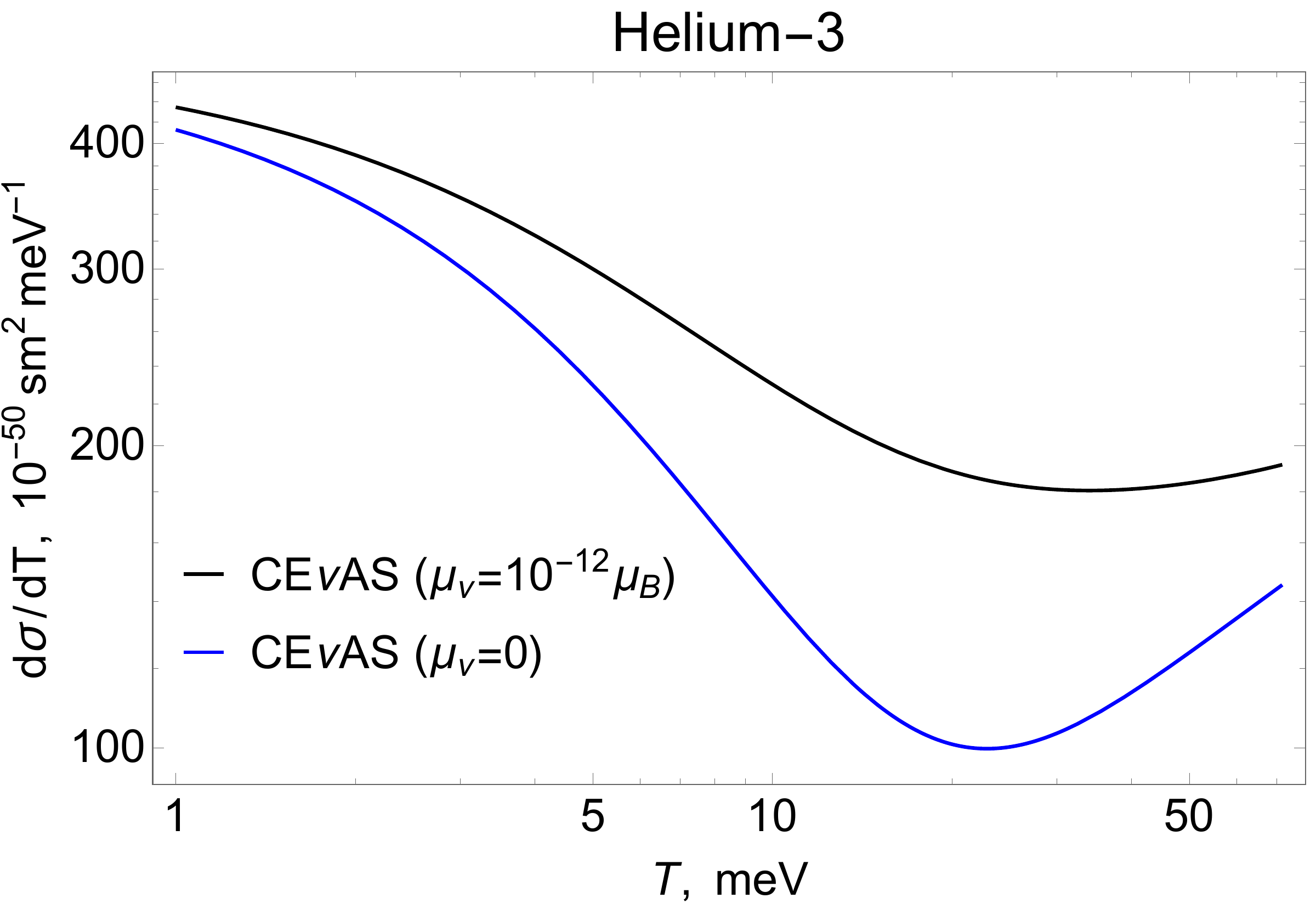}
        \hfill
    \includegraphics[width=0.48\linewidth]{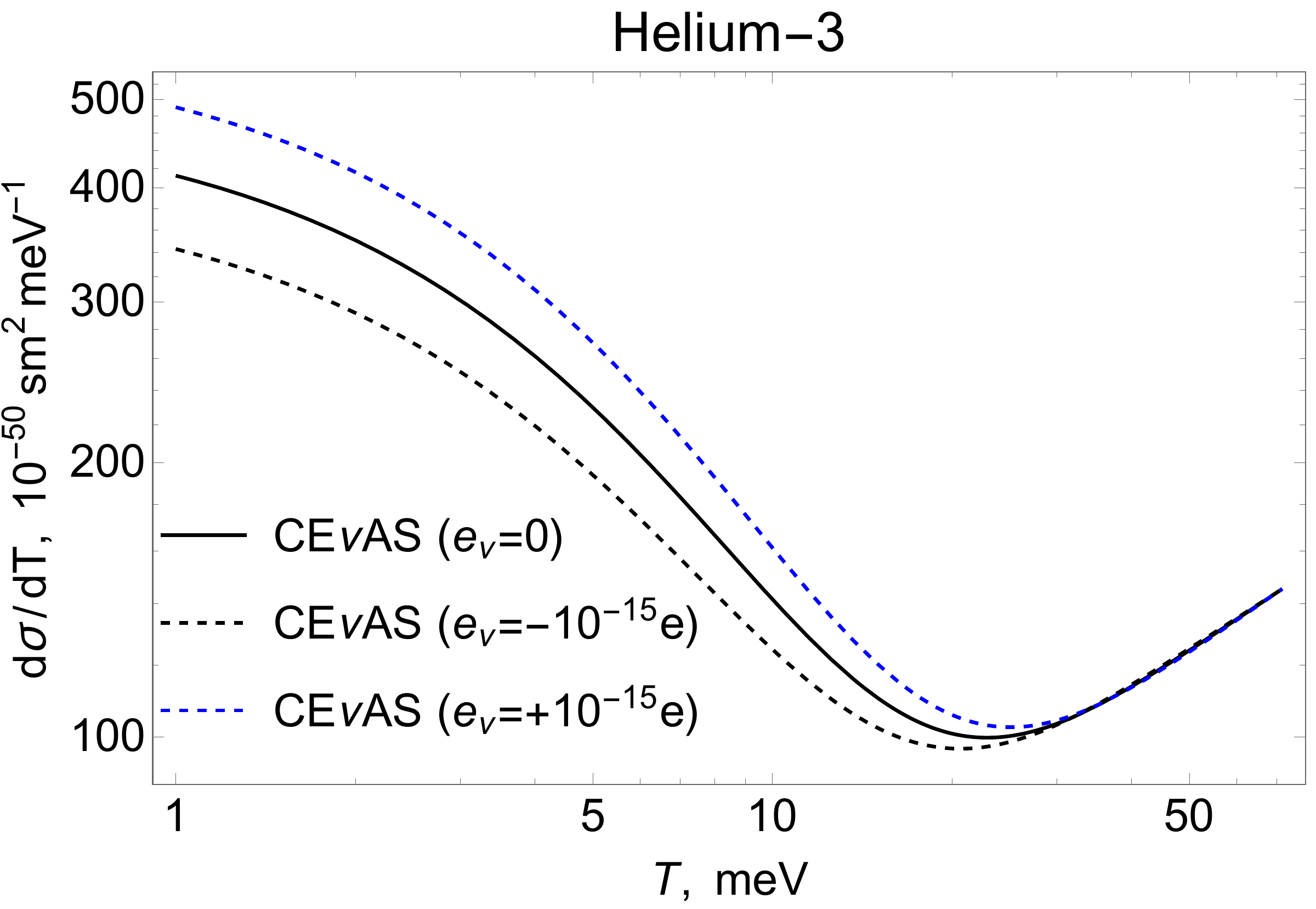}\\
   \vspace{3mm}
        \includegraphics[width=0.48\linewidth]{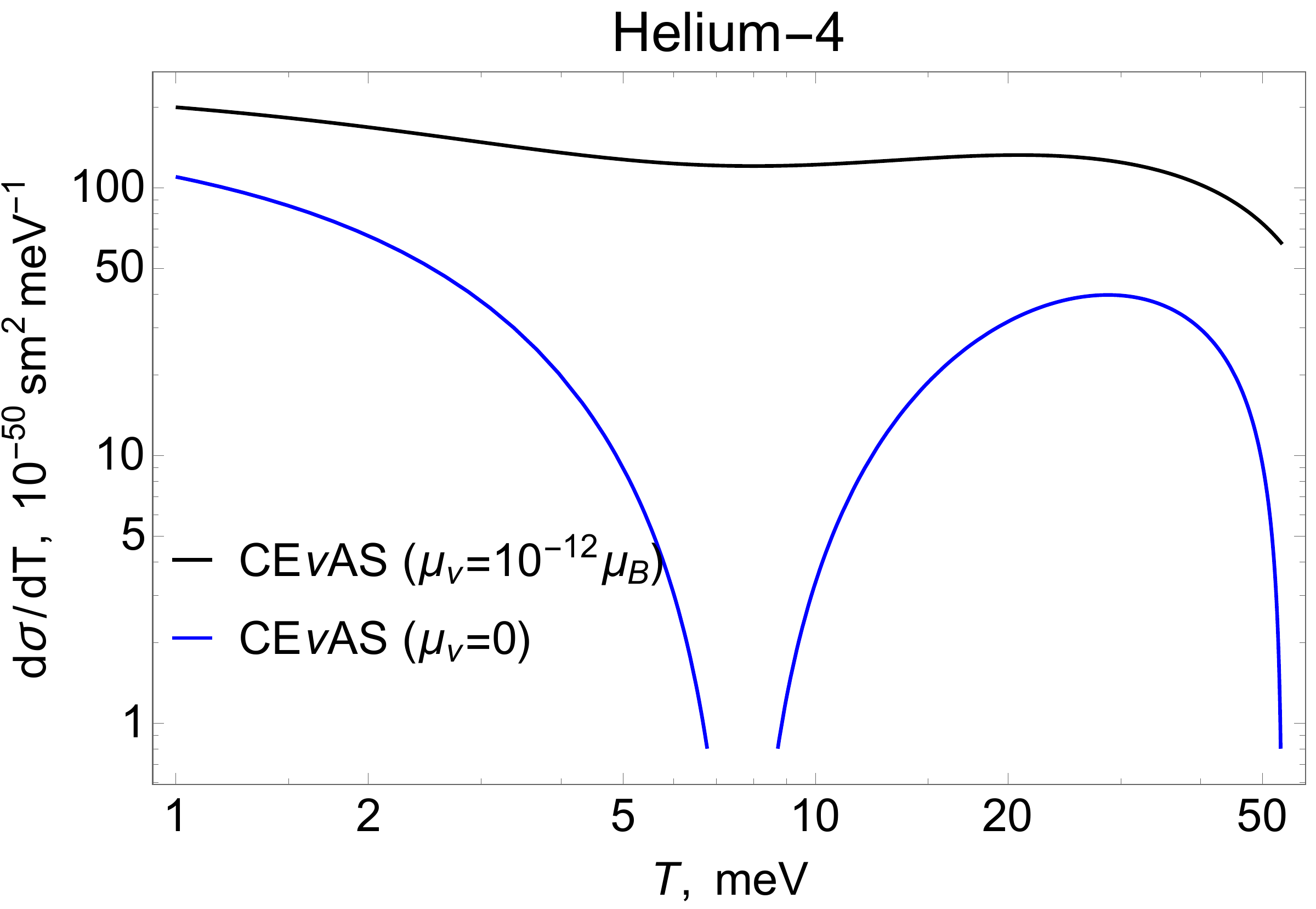}
        \hfill
    \includegraphics[width=0.48\linewidth]{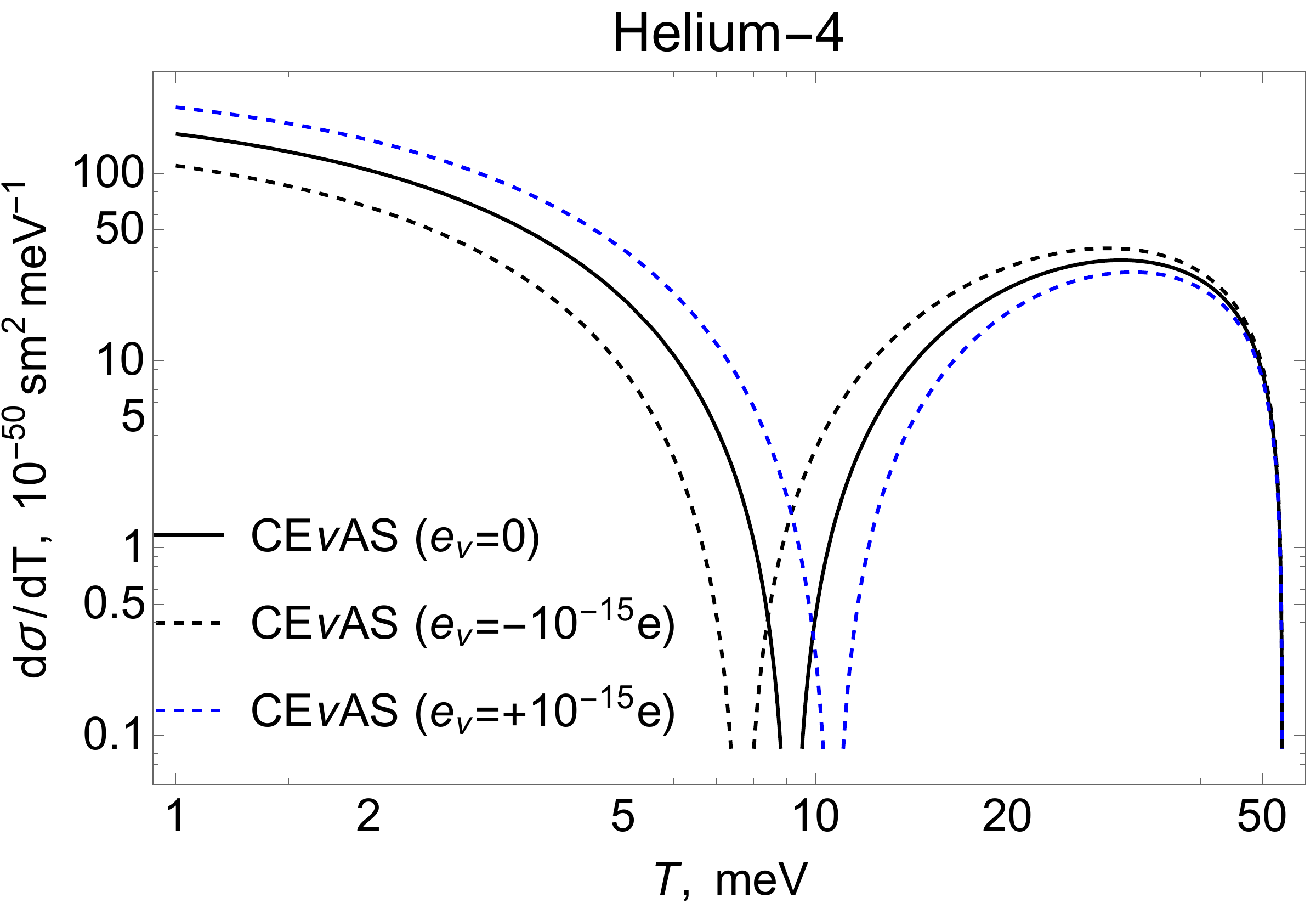}\\
 %  \vspace{3mm}
%        \includegraphics[width=0.46\linewidth]{figs/C12_mm_2.1.pdf}
%        \hfill
%    \includegraphics[width=0.46\linewidth]{figs/C12_e_2.1.pdf}
    \caption{Differential cross sections for the CEvAS processes within Standard Model ($e_\nu=0$ and $\mu_\nu=0$ ) and with account for the neutrino millicharge ($e_\nu=\pm10^{-15}e$) and magnetic moment ($\mu_\nu=10^{-12}\mu_B$).}
    \label{fig:my_label}
\end{figure}
The neutrino magnetic moment contribution is
\begin{equation}\frac{d\sigma^{(\mu_\nu)}}{dT} = \frac{\pi\alpha^2 Z^2}{m_e^2} |\mu_\nu|^2 \left(\frac{1}{T} - \frac{1}{E_\nu} \right) \left[1 - F_{\rm el}(q^2)\right]^2,
\end{equation}
where the neutrino magnetic moment $\mu_\nu$ is in units of $\mu_B$. In contrast to the case of neutrino millicharge, the neutrino magnetic moment interaction flips the neutrino helicity, and therefore it does not interfere with the weak interaction channel.

In Fig.~\ref{fig:my_label} we present the numerical results for the differential cross section~(\ref{cr_sec}) in the case of an electron antineutrino with $E_\nu=10$ keV that is typical for the tritium neutrino source. It can be seen that the atomic recoil spectra in CE$\nu$AS processes on the H, $^2$H, $^3$He, and $^4$He %and $^{12}$C
atomic systems are very sensitive to the neutrino millicharge and magnetic moment. Measuring these spectra may allow us to test the $e_\nu$ and $\mu_\nu$ values at a level of $10^{-15}e$ and $10^{-12}\mu_B$, respectively, or even below that level.

The obtained results will be used in the search for the electromagnetic properties of neutrinos in the experiment involving an intense tritium neutrino source and a superfluid $^4$He target. This experiment is currently being prepared in the framework of the research program of the National Center for Physics and Mathematics in Sarov, Russia.
%

%
%\section{Conclusions}
%
%We accounted for the neutrino millicharge and magnetic moment in the theory of CE$\nu$AS. It is shown that the atomic recoil spectra in CE$\nu$AS processes on the H, $^2$H, $^3$He, and $^4$He %and $^{12}$C
%atomic systems are very sensitive to the neutrino millicharge and magnetic moment. Measuring these spectra may allow us to test the $e_\nu$ and $\mu_\nu$ values at a level of $10^{-15}e$ and $10^{-12}\mu_B$, respectively, or even below that level. The obtained results will be used in the search for the electromagnetic properties of neutrinos in the experiment involving an intense tritium neutrino source and a superfluid $^4$He target. This experiment is currently being prepared in the framework of the research program of the National Center for Physics and Mathematics in Sarov, Russia (parallel talk, presentation \#775).

%
\section*{Acknowledgments}
The work is supported by the Russian Science Foundation under grant No. 22-22-00384. G.D. acknowledges the support from the National Center for Physics and Mathematics (Project ``Study of coherent elastic neutrino-atom and -nucleus scattering and neutrino electromagnetic properties using a high-intensity tritium neutrino source'').

\end{document}